\documentclass[twocolumn,pre,superscriptaddress,floatfix,amsmath,amssymb,aps]{revtex4-2}
\pdfoutput=1

\usepackage[utf8]{inputenc}
\usepackage{natbib}
\usepackage{array}
\usepackage{booktabs}
\usepackage{graphicx}
\usepackage{paralist}
\usepackage{color,soul}
\usepackage[colorlinks=true, allcolors=blue]{hyperref}
\usepackage{mathtools}
\usepackage{amsmath}
\usepackage{dcolumn}
\usepackage{amssymb}
\usepackage{gensymb}
\usepackage{mathdesign}
\usepackage{siunitx}
\usepackage{accents}
\usepackage{bm}
\usepackage{dcolumn}

\begin{document}

\title{Fractal dimensions of jammed packings with power-law particle size distributions in two and three dimensions}
\author{Joseph M. Monti}
\affiliation{Sandia National Laboratories, Albuquerque, NM 87185, USA}
\author{Ishan Srivastava}
\affiliation{Center for Computational Sciences and Engineering, Lawrence Berkeley National Laboratory, Berkeley, California 94720, USA}
\author{Leonardo E. Silbert}
\affiliation{School of Math, Science, and Engineering, Central New Mexico Community College, Albuquerque, New Mexico 87106, USA}
\author{Jeremy B. Lechman}
\affiliation{Sandia National Laboratories, Albuquerque, NM 87185, USA}
\author{Gary S. Grest}
\affiliation{Sandia National Laboratories, Albuquerque, NM 87185, USA}
\date{\today}

\begin{abstract}
Static structure factors are computed for large-scale, mechanically stable, jammed packings of frictionless spheres (three dimensions) and disks (two dimensions) with broad, power-law size dispersity characterized by the exponent $-\beta$. 
The static structure factor exhibits diverging power-law behavior for small wavenumbers, allowing us to identify a structural fractal dimension, $d_f$. 
In three dimensions, $d_f \approx 2.0$ for $2.5 \le \beta \le 3.8 $, such that each of the structure factors can be collapsed onto a universal curve. 
In two dimensions, we instead find $1.0 \lesssim d_f \lesssim 1.34 $ for $2.1 \le \beta \le 2.9 $. 
Furthermore, we show that the fractal behavior persists when rattler particles are removed, indicating that the long wavelength structural properties of the packings are controlled by the large particle backbone conferring mechanical rigidity to the system.
A numerical scheme for computing structure factors for triclinic unit cells is presented and employed to analyze the jammed packings.
\end{abstract}

\maketitle

An underlying theme in the study of granular materials is the ability to determine the structural arrangement of the grains that constitute a static, mechanically stable, particle packing. 
In regular thermal systems, the small wavenumber limit of the static structure factor relates to the mechanical properties of the system \cite{hansenmcdonald}.  
For disordered and amorphous jammed packings of frictionless and monodisperse spheres, the small wavenumber ($q$) behavior of the structure factor, $S(q)$, expresses a suppression of density fluctuations at large length scales termed hyperuniformity, i.e., $S(q) \sim q$ \cite{donev2005,silbert2009}. 
When particle size dispersity is introduced into the packing, the observed hyperuniformity is lost, even for a bidisperse packing of spheres \cite{xu2010}. 
Though this behavior can be reconciled through an appropriate combination of the partial structure factor contributions due to the different particle size species \cite{hansenmcdonald,berthier2011,weeks2010}, the procedure is only manageable over a restricted range of dispersity. 
It is not uncommon for colloidal suspensions to exhibit fractal properties~\cite{sorensen2001}: namely, that for small to intermediate wavenumbers, the structure factor diverges as $S(q) \sim q^{-d_{f}}$, which defines the (structural) fractal dimension, $d_{f}$. 
What remains unclear is the extent to which broad, continuously-distributed dispersity influences the structure of sphere packings.
We address this question in this Letter using power-law size distributions of spherical particles.

For power-law size distributions, the number of particles with diameters $D$ in a small increment between $D$ and $D+\Delta D$ is $N(D) \Delta D \propto D^{-\beta} \Delta D$, where $\beta$ is the distribution exponent. 
Most studies of the packings of $d$-dimensional spheres with a power-law size distribution have focused on the conditions under which one can achieve full packing. 
The most well-known example is the Apollonian packing, in which space is filled with $d$-dimensional spheres by iteratively fitting the next sphere into the largest available void.  
As each new particle is jammed by its neighbors, the packing is inherently mechanically stable~\cite{anishchik1995}. 
The resulting packings have a fractal dimension $d_f^{\rm Ap} = 1.3057...$ in 2 dimensions ($2d$) and $2.4739...$ in 3 dimensions ($3d$)~\cite{borkovec1994,anishchik1995,varrato2011}. 
~\citet{aste1996} conjectured that the full packing is possible for power-law distributions with $\beta$ between $d_f^{\rm Ap}+1$ and $d+1$, while ~\citet{botet2021} showed that size distributions with $3.8 \leq \beta < 4$ are space filling in $3d$.
Several randomized packing strategies have been employed that produce power-law particle assemblies, including the random sequential addition algorithm (RSA)~\cite{cherny2023,torquato2006} that iteratively fills space with progressively smaller spheres, and packing-limited growth strategies~\cite{andrienko1994,dodds2002} that nucleate and swell particles until jamming.
These packings have fractal-like structure with $d_f\equiv\beta-1$.

In this Letter, we take a more physically-motivated approach to generate packings using discrete element method (DEM) simulations to compress a dilute assembly of power-law distributed particles until jamming.
Using large-scale simulations with particle size ratios of up to 300 in $2d$ and up to 200 in $3d$, we show that the fractal dimensions of these packings computed from $S(q)$ are $d_f \approx 2.0$, independent of $\beta$ in $3d$.  
In $2d$, $d_f \sim 1$ for small $\beta$ and saturates at $d_f \approx 1.34$ for larger $\beta$.
This is in contrast to results for the RSA packing method, for which~\citet{cherny2023} found $d_f=\beta-1$ in $1d$ and $2d$.

In the DEM simulations, particles interact via frictionless, damped, purely repulsive Hookean springs.
The normal force $\mathbf{F}_{\rm n}$ between contacting particles $i$ and $j$ separated by $\mathbf{r}_{ij} = \mathbf{r}_i - \mathbf{r}_j$ is~\cite{cundall1979,silbert2001}
{\color{black}
\begin{equation}
    \mathbf{F}_{\rm n} = k_{\rm n}\delta\frac{\mathbf{r}_{ij}}{|\mathbf{r}_{ij}|}- M_{\rm eff}\gamma_{\rm n}\mathbf{v}_{\rm n},
    \label{eq:force}
\end{equation}
}
where $k_{\rm n}$ is the spring stiffness {\color{black} set equal to unity} and $\delta = (D_i + D_j)/2 - |\mathbf{r}_{ij}|$ is the overlap in terms of the diameters $D_i$ and $D_j$.
The second term on the right-hand side penalizes relative normal velocity $\mathbf{v}_{\rm n}$ with strength proportional to the effective particle mass $M_{\rm eff} = M_iM_j/(M_i+M_j)$ and a damping coefficient $\gamma_{\rm n} $ {\color{black} set equal to 0.5}.
Particle mass densities are set to unity so that $M_i = \pi D_i^3/6$.

Particle sizes fall in the range $\sigma\le D \le \lambda\sigma$, where $\lambda$ denotes the maximum size ratio and $\sigma$ is the diameter of the smallest particle, which is set to unity and is used to non-dimensionalize $q$.
Systems are required to have at least ten particles with diameters larger than $0.95\lambda\sigma$ in $3d$ and fifty such particles in $2d$, meaning that the total number of particles, $N$, depends upon both $\lambda$ and $\beta$---see Table~\ref{tab:table1} for system details. 
{\color{black} $\beta$ is selected from $2.5\le \beta \le 3.8$ in $3d$ and $2.1 \le \beta \le 2.9$ in $2d$;  these ranges separately include the Apollonian packing exponents $\beta^{\rm Ap} \approx 2.31$ ($2d$) and $\beta^{\rm Ap} \approx 3.47$ ($3d$).
In $3d$, particle counts become intractable for $\beta \rightarrow 4.0$ with large $\lambda$.
Simulated $\lambda$ values vary depending on $d$, $\beta$, and computational limits to obtain sufficient scaling regimes to reliably extract the fractal dimension $d_f$, or to isolate the role of $\lambda$ specifically.}

Packings are generated with the GRANULAR package in LAMMPS~\cite{thompson2022} using a constant-pressure protocol~\cite{santos2020,srivastava2021}.
Our simulations use an efficient particle-size-based neighbor binning algorithm~\cite{ogarko2012,krijgsman2014,stratford2018,shire2021} that has been used to study both bidisperse and power-law distributed systems~\cite{srivastava2021,monti2022a,monti2022b}.
The simulation box is periodic, initially cubic (square in $2d$), and dilute with particles placed randomly without overlaps.
Packing proceeds by imposing an isotropic applied pressure tensor, $\mathbf{P}_{\rm a}$, with diagonal components set to a constant, $p_{\rm a}$, and off-diagonal components set to zero, and stops when the internal pressure tensor matches $\mathbf{P}_{\rm a}$ and the kinetic energy per particle is small.
The simulation cell deforms from cubic to slightly triclinic in order to relax off-diagonal stress components---we introduce a mathematical procedure for calculating $S(q)$ for triclinic unit cells in the Appendix.
Simulations conducted in $2d$ constrain motion to the $x-y$ plane and use a $2d$ applied pressure tensor but are otherwise identical to the $3d$ simulations.
Here, we combine results using both $p_{\rm a} = 10^{-6} k_{\rm n}/\sigma$ and $10^{-4}k_{\rm n}/\sigma$ without significantly affecting structural measures as both pressures produce packings in the small overlap limit~\cite{santos2020}.

\begin{table}
\caption{Particle-size distribution parameters and structural properties: {\color{black} physical dimension $d$, size distribution exponent $\beta$, maximum particle size ratio $\lambda$, representativity ratio $L/\lambda\sigma$, total particle count $N$, packing volume fraction $\rho$, non-rattler particle packing volume fraction $\rho_{NR}$, rattler particle number fraction $\phi_{R}$, and packing fractal dimension $d_f$. In $3d$, the fractal range is insufficient to obtain $d_f$ for $\beta = 3.3$, $\lambda = 32$.}}
\begin{tabular}{|c|c|c|c|c|c|c|c|c|}
  \hline
   $d$ &$\beta$ &$\lambda$ & ${\color{black} L/\lambda\sigma} $ &$N$ & $\rho$ & $\rho_{NR}$ & $\phi_R$  & $d_f$ \\
  \hline
  2 & 2.1  & 300 & {\color{black} 29.8} &459,651 & 0.932 & 0.873 & 0.824 & 0.97$\pm$0.01  \\
  2 & 2.3  & 200 & {\color{black} 46.9} &1,430,164 & {\color{black} 0.943} & {\color{black} 0.896} & {\color{black} 0.620} &  1.14$\pm$0.01  \\
  2 & 2.5  & 200 & {\color{black} 63.9} &5,020,131 & 0.950 & 0.918 & 0.420 & 1.28$\pm$0.01 \\
  2 & 2.7  & 100 & {\color{black} 62.7} &2,762,800 & 0.934 & 0.907 & 0.239 & 1.33$\pm$0.01  \\
  2 & 2.9  & 100 & {\color{black} 54.0} &3,089,316 & 0.925 & 0.898 & 0.179 & 1.34$\pm$0.01 \\
  \hline
  3 & 2.5  & 200 & {\color{black} 8.8} &2,848,307  & 0.763 & 0.699 & 0.995 & 2.05$\pm$0.05\\
  3 & 2.75 & 150 & {\color{black} 6.7} &2,060,317 & 0.772 & 0.669 & 0.996 &2.0$\pm$0.02 \\
  3 & 3.0  & 150 & {\color{black} 4.9} &2,089,645 & 0.812 & 0.691 & 0.997 &2.02$\pm$0.01 \\
  3 & 3.3  & 100 & {\color{black} 5.4} &3,187,515 & 0.857 & 0.745 & 0.977 &1.97$\pm$0.01 \\
  3 & 3.3  & 50  & {\color{black} 9.2} &{\color{black} 3,260,524 }& {\color{black} 0.832 } & {\color{black} 0.744 } & {\color{black} 0.862} & 1.97$\pm$0.01  \\
  3& 3.3 & 50  & {\color{black} 5.4 }& 652,106 & 0.839 & 0.746 & 0.911 &1.97$\pm$0.01 \\
  3 & 3.3  & 32  & {\color{black} 12.9} &{\color{black} 3,259,156 } & {\color{black} 0.814 } & {\color{black} 0.737} & {\color{black} 0.772} & -- \\
  3 & 3.3 & 32 & {\color{black} 5.4 } & 232,797 & 0.818 & 0.738 & 0.810 & -- \\
  3 & 3.6 & 50  & {\color{black} 6.1} &1,851,063 & 0.839 & 0.797 & 0.527 &2.0$\pm$0.01  \\
  3 & 3.8 & 50  & {\color{black} 6.8} &3,739,236&  0.822 & 0.791 & 0.338 & 2.0$\pm$0.01 \\
    \hline
\end{tabular}
\label{tab:table1}
\end{table}

\begin{figure}[ht]
\includegraphics[width=1.0\linewidth]{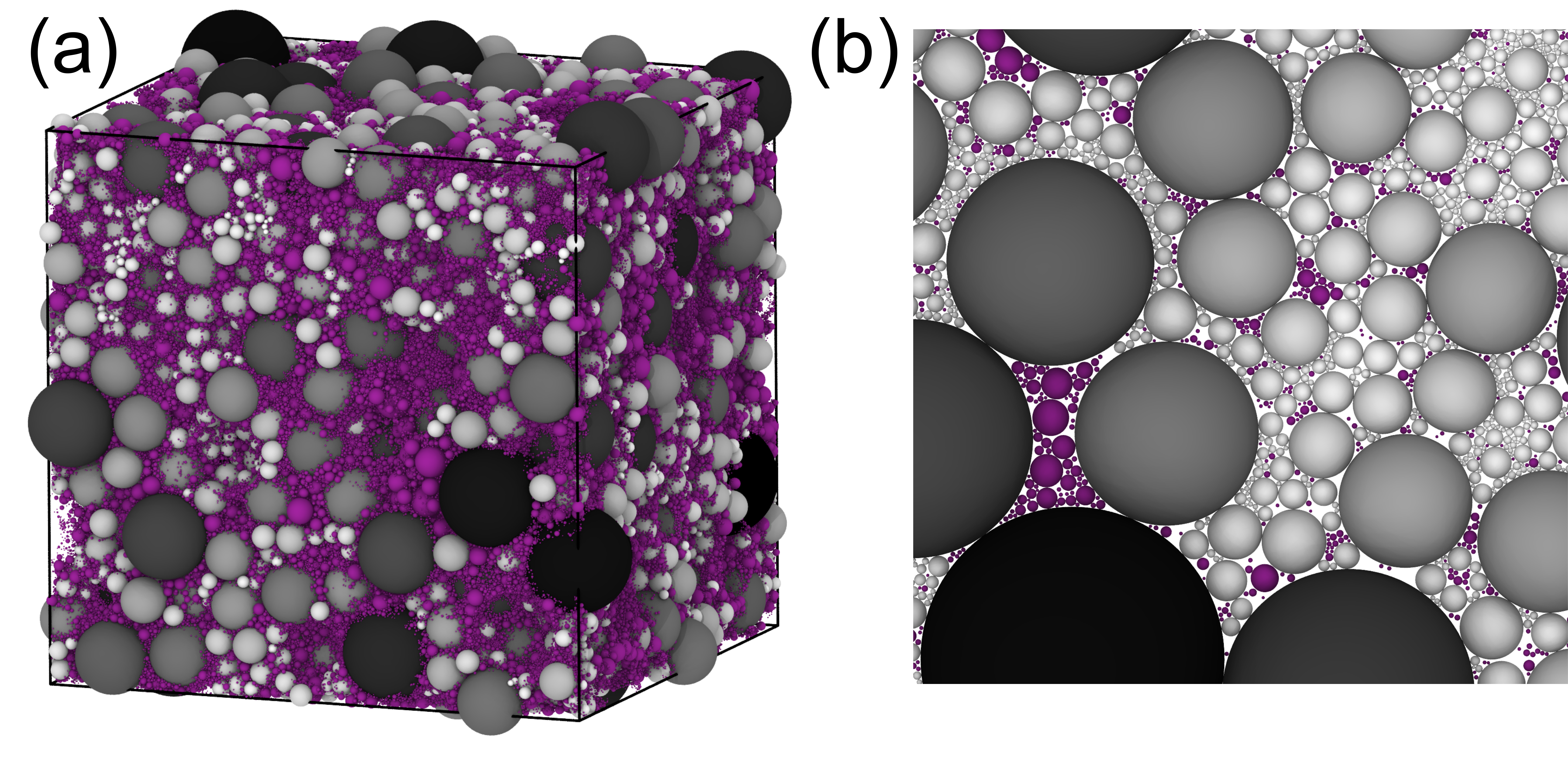}
\caption{OVITO snapshots~\cite{ovito} of jammed systems with maximum sphere diameter $\lambda\sigma = 100$ in $3d$ with $\beta=3.0$ (a) and $2d$ with $\beta = 2.3$ (b). Non-rattler particles are shaded light to dark to indicate increasing diameter. Rattler particles are shown in purple, with fraction of rattlers $\phi_R = 0.993$ in $3d$ and 0.497 in $2d$.}
\label{fig:snapshots}
\end{figure}

{\color{black} In this Letter, we report results of individual simulations for each set of parameters.}
{\color{black} Table~\ref{tab:table1} includes the ratio of the equivalent simulation cell length, $L$, to the largest physical length scale $\lambda\sigma$, where $L \equiv V^{1/d}$ in terms of the compacted simulation cell volume $V$.}
{\color{black} With $L/\lambda\sigma > 1$, simulations are self-averaging because the cell is composed of $\sim (L/\lambda\sigma)^d$ small volume replicas.}
{\color{black} Furthermore, we will show that the fractal properties of the packing are independent of the quality of the packing, i.e., when mechanical equilibrium is not exactly (numerically) satisfied, so long as the largest particles are nearly at rest.}

Recent studies employing DEM simulations to generate packings of systems composed of power-law size distributions~\cite{monti2022a} or power-law cumulative mass distributions~\cite{estrada2016,oquendo2020,oquendo2021,oquendo2022} have demonstrated that the packing volume fraction (area fraction in $2d$), $\rho$, strongly depends on both the power-law exponent characterizing the distribution and the distribution span.
These simulations showed, in 2$d$~\cite{estrada2016} and $3d$~\cite{oquendo2020,oquendo2021,oquendo2022,monti2022a}, that $\rho$ reaches a maximum for {\color{black} distributions with $d < \beta < d+1$}, and that increasing $\lambda$ produces denser packings with other parameters held constant, similar to particle insertion techniques like the Apollonian packing.
This dependence of $\rho$ on structural parameters is also indicated in Table~\ref{tab:table1}. 
~\citet{monti2022a} provided evidence that the distribution of contacts between particles of disparate sizes shifts over the same range of exponents: for $\beta$ approaching $d+1$, the largest particles in the packing tend to be stabilized by a saturation of small particle neighbors, while for $\beta$ approaching $d$, the backbone force network supporting the applied pressure is composed primarily of the largest particles.
Indeed, in the latter case, only these largest particles are mechanically stable, i.e., they are held in place by sufficiently many neighboring stable particles.
Without performing systematic structural analysis of the packings, however, it is unclear if such changes in particle connectivity with distribution exponent are reflected in the position correlation functions.

Packings composed of power-law or power-law-like size distributions generally possess an abundance of mechanically unstable particles, termed rattlers.
Non-rattler particles are identified through an iterative procedure~\cite{donev2004} by isolating those with at least $d$ stable neighbors.
The fraction of all particles that are rattlers is denoted $\phi_R$ and the corresponding values are listed in Table~\ref{tab:table1}.
Similar to the particle volume fraction, $\phi_R$ depends on both $\beta$ and $\lambda$, with the largest values found for large $\lambda$ and $\beta \lesssim d$.
Figure~\ref{fig:snapshots} shows snapshots of prototypical systems in $3d$ (full system) and $2d$ (partial system) with rattlers indicated in color, with $\phi_R = 0.993$ and 0.497, respectively.
Note that while the non-rattler particles shown in Fig.~\ref{fig:snapshots}(a) number fewer than $10^4$, most of the total particle volume is contained in these particles.
Table~\ref{tab:table1} also enumerates the particle volume fraction contributed exclusively by non-rattler particles, $\rho_{NR}$.
Rattler particle positions are somewhat arbitrary depending on their size relative to the pore space they inhabit and by how many other rattlers are nearby; several examples of pockets of rattlers are visible in Fig.~\ref{fig:snapshots}(b).
Because rattlers generally originate from the small particle limit of the size distribution~\cite{monti2022a}, noisiness associated with their positions has most significant effect at small length scales, and we will show that the estimate of the packing fractal dimension is unaffected by removing them.

\begin{figure}
\includegraphics[width=0.8\linewidth]{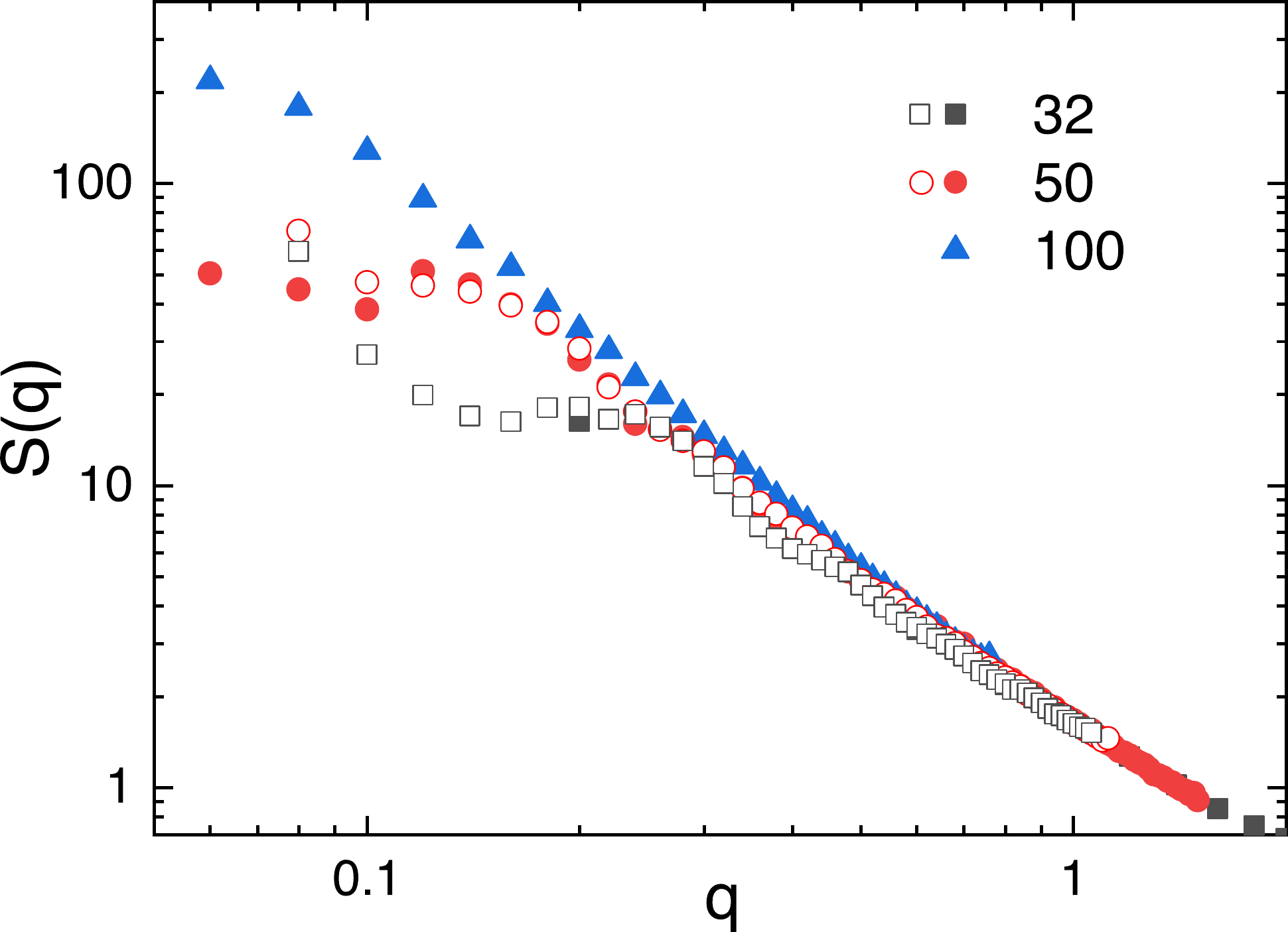}
\caption{Static structure factors $S(q)$ for $3d$ packings with $\beta=3.3$ and the indicated maximum particle size ratios $\lambda$. {\color{black} Open and filled symbols respectively distinguish systems with greater and lesser particle numbers, $N$, listed in Table~\ref{tab:table1}.}}
\label{fig:S(q)-3.3}
\end{figure}

\begin{figure}
\includegraphics[width=0.8\linewidth]{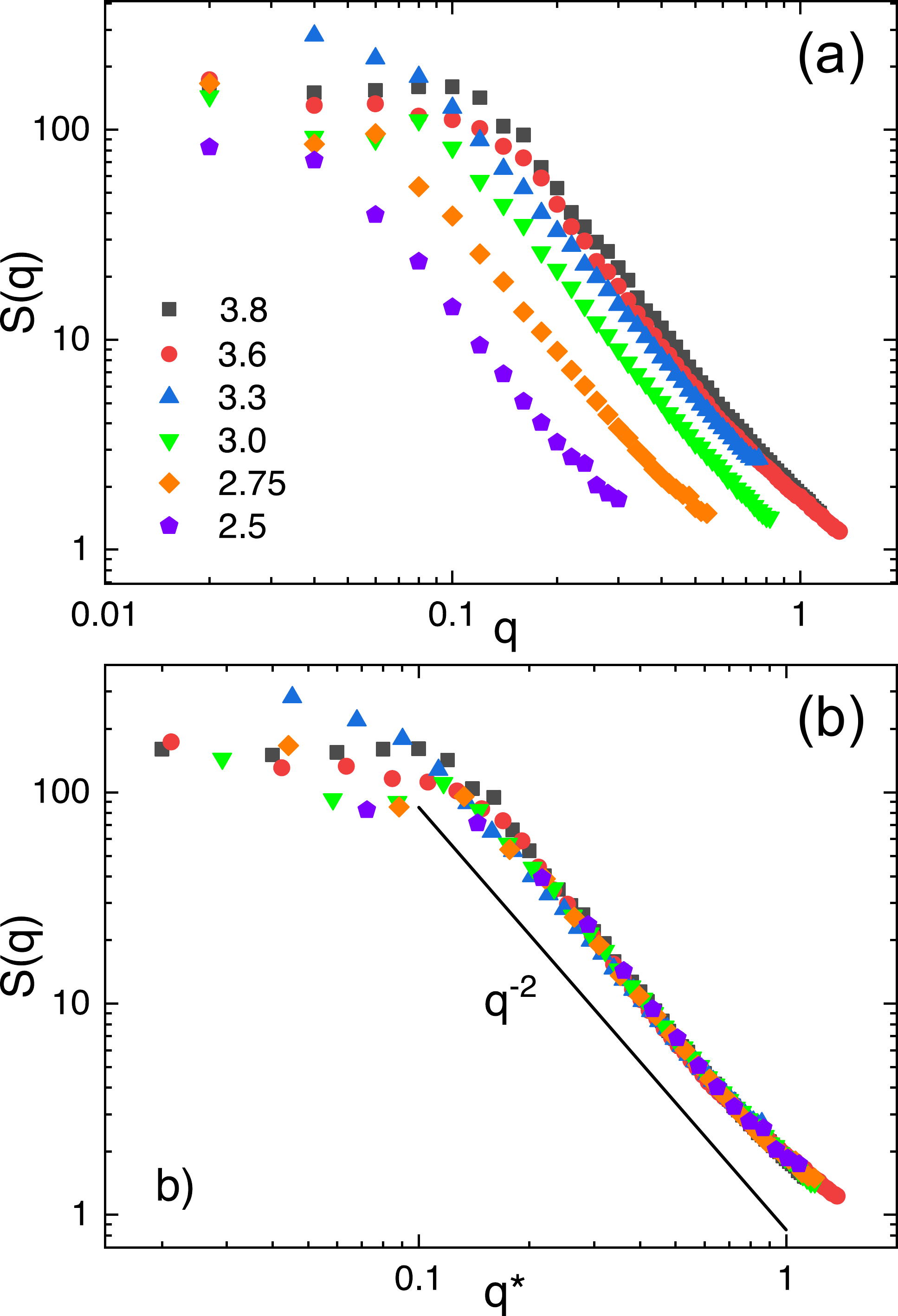}  
\caption{(a) $S(q)$ in  $3d$ for the indicated $\beta$, {\color{black} with $\lambda = 200\ (\beta = 2.5)$, $\lambda = 150\ (\beta = 2.75 \text{ and } 3.0)$, $\lambda = 100\ (\beta = 3.3)$, and $\lambda = 50\ (\beta = 3.6 \text{ and } 3.8)$.} (b) The same data, plotted as $q^* = q\alpha(\beta)$ with $\alpha(\beta) = \lbrace 1.00, 1.06, 1.13, 1.46, 2.21, 3.61 \rbrace$ for $\beta = \lbrace 3.8, 3.6, 3.3, 3.0, 2.75, 2.5 \rbrace$. }
\label{fig:S(q)_3d}
\end{figure}

The static structure factor $S(q)$ is shown in Fig.~\ref{fig:S(q)-3.3} for $\beta=3.3$ in $3d$ for three values of $\lambda$, focusing on small to intermediate $q$ values.
$S(q)$ rolls over to a constant value below $q \sim 2\pi/\lambda$, indicating that the packing structure is uniform over the corresponding (large) length scales in this regime, which extends down to $q \sim 2\pi\sigma/L$. 
For intermediate $q$, $S(q)$ exhibits power-law, i.e., fractal, scaling for which $S(q) \sim q^{-d_f}$ in terms of the fractal dimension.
{\color{black} The power-law best fit from least-squares fitting over $0.08 \leq q \leq 0.8$ in Fig.~\ref{fig:S(q)-3.3} gives $d_f = 1.97 \pm 0.01$ for the broadest fractal regime, for $\lambda = 100$.
The figure shows that the fractal regime grows as $\lambda$ increases, because $\lambda$ acts as the cutoff length scale of fractal behavior---this must be the case, as $\lambda\sigma$ is the only large physical length scale that falls in between $\sigma$ and $L$.
The presence of the rollover in $S(q)$ below $q \sim 2\pi/\lambda$ signifies that increasing system size at constant $\lambda$ broadens the range of homogeneous behavior at low $q$, but has no effect on the fractal regime.
Figure~\ref{fig:S(q)-3.3} demonstrates that the rollover $q$ value is independent of $N$ by varying $N$ for fixed $\lambda = 32$ and 50 (open and filled symbols).
Note that $N$ values are comparable for the open symbols and the $\lambda = 100$ data. }

Results for $S(q)$ for $3d$ packings comprised of power-law size distributions with a range of exponents are plotted in Fig.~\ref{fig:S(q)_3d}(a). 
Values of $\lambda$ for each system were chosen to give a consistent power-law scaling regime for estimating $d_f$ from $S(q)$. 
{\color{black} Least-squares fitting implies that the fractal dimension is independent of $\beta$ in $3d$ with value $d_f \approx 2.0$, as suggested by the solid line drawn in Fig~\ref{fig:S(q)_3d}(b)}.
Figure~\ref{fig:S(q)_3d}(b) shows that the data collapse if $q$ is scaled as $q^*=\alpha(\beta)q$, where $\alpha(\beta)$ are suitable (length-) scaling prefactors, with the arbitrary convention $\alpha(\beta = 3.8) = 1.0$; the associated $\alpha(\beta)$ values are given in the figure caption.

The independence of $d_f$ from $\beta$ in $3d$ is surprising given that every other packing quantity we computed depends on the input distribution; see Table~\ref{tab:table1} and Ref.~\cite{monti2022a}.
Moreover, this result is at odds with a separate, conventional definition of $d_f$ defined for collections of particles~\cite{turcotte1986}, which is obtained from extracting the exponent characterizing the power-law mass or size distribution without regard to particle configurations.
This definition of fractal dimension implies $d_f(\beta) = \beta - 1$ for power-law size distributions, obviously distinct from our result, with the caveat that our simulations assume a random configuration of frictionless particles compacted from a well-mixed, dilute state.

\begin{figure}
\includegraphics[width=0.8\linewidth]{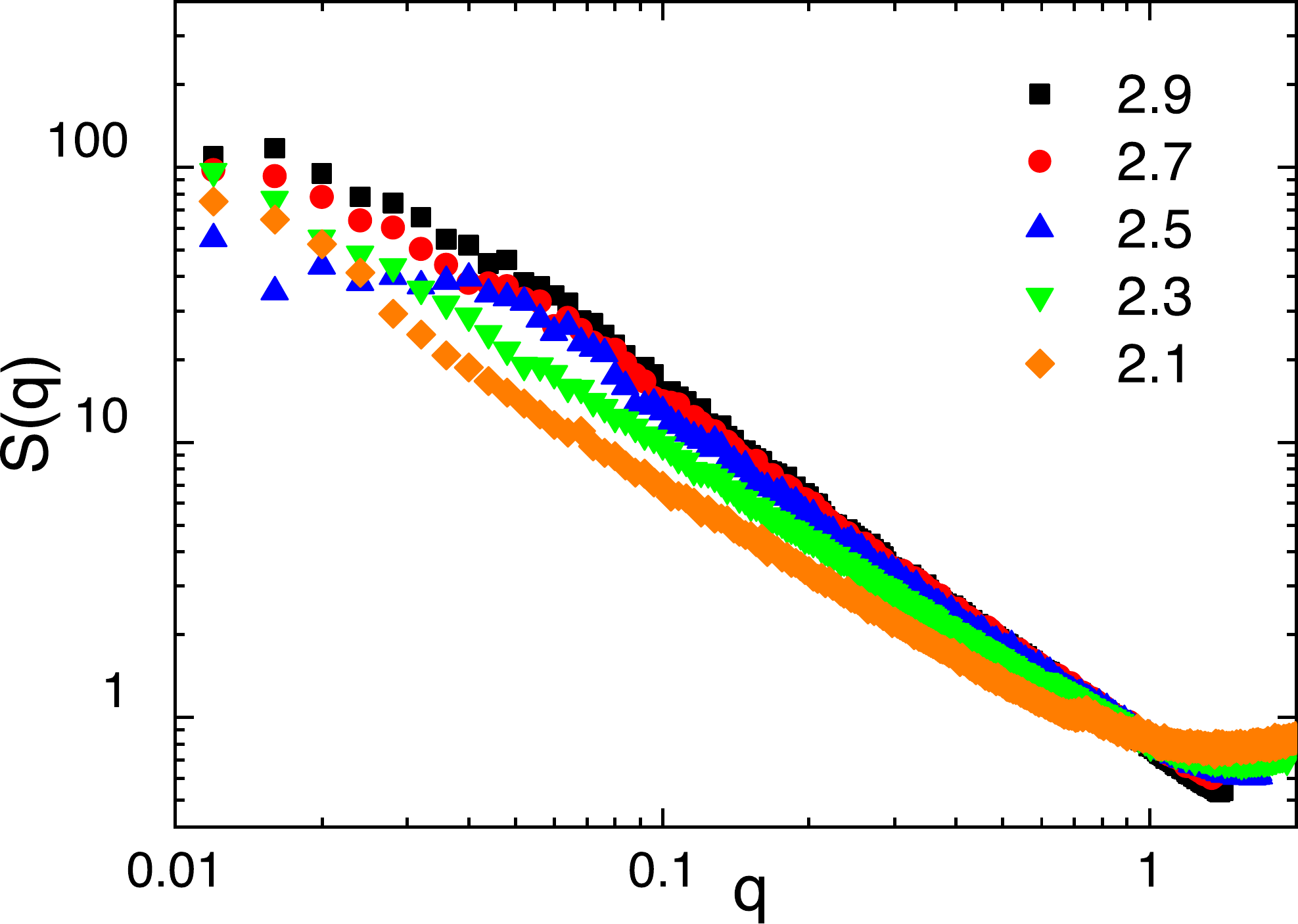}
\caption{$S(q)$ in  $2d$ for the indicated $\beta$, {\color{black} with $\lambda = 300\ (\beta = 2.1)$, $\lambda = 200\ (\beta = 2.3 \text{ and } 2.5 )$, and $\lambda = 100\ (\beta = 2.7 \text{ and } 2.9)$.}}
\label{fig:S(q)_2d}
\end{figure}

Similarly to $S(q)$ computed for $3d$ packings, $S(q)$ for a range of $\beta$ in $2d$ also exhibit power-law scaling regimes.
Our results for $2d$ $S(q)$ are plotted in Fig.~\ref{fig:S(q)_2d}.
Unlike the $3d$ case, conversely, the estimated fractal dimensions in $2d$ show some dependence on $\beta$.
We find that $d_f$ increases from $0.97\pm 0.01$ for the lowest value $\beta = 2.1$ to $1.27\pm 0.01$ for $\beta=2.5$ before appearing to saturate at approximately $4/3$ for the largest values of $\beta$. 
Note that the largest $\lambda$ simulated was $ 300$ in $2d$, and consequently the power-law regimes are broader in $2d$ compared to $3d$. 

One expects that $2d$ packings should be more sensitive to DEM packing protocol than $3d$ packings in that particles cannot be squeezed through constrictions formed between existing contacts, as they can in $3d$.
This effect produces pockets of trapped but mobile small particles and becomes more pronounced for distributions with $\beta \lesssim 2.5$, for which contacts between large particles are more frequent; see Fig.~\ref{fig:snapshots}(b), for example.
In $3d$, particles smaller than nearby constrictions may be able to escape between pores, alleviating the effect to an extent.
Particle insertion techniques have no analogy to this phenomenon because particles are static and insertion only depends on the local pore size.

A recent model and numerical work conducted by~\citet{cherny2023} using a RSA algorithm for $d$-dimensional spheres (in $1d$ and $2d$) produced assemblies with $d_f = \beta - 1$, as computed from static structure factor calculations, in addition to the associated mass-radius and pair-distribution functions.
The authors also presented approximations to the effect that the $1d$ and $2d$ results could be extended to higher dimensions without changing the conclusion regarding $d_f$.
The specific $2d$ exemplar considered in Ref.~\cite{cherny2023} was $\beta -1 = d_f = 1.4$, which differs by roughly 0.2 from an estimate from our DEM-obtained results (i.e., using the midpoint of the $\beta = 2.3$ and 2.5 results in Table~\ref{tab:table1}).
However, the difference in $\beta$-dependence between our observations in $2d$ and especially $3d$, and the findings of~\citet{cherny2023} is irreconcilable, implying that the packing protocol fundamentally differentiates the final particle structures.

\begin{figure}
\includegraphics[width=0.8\linewidth]{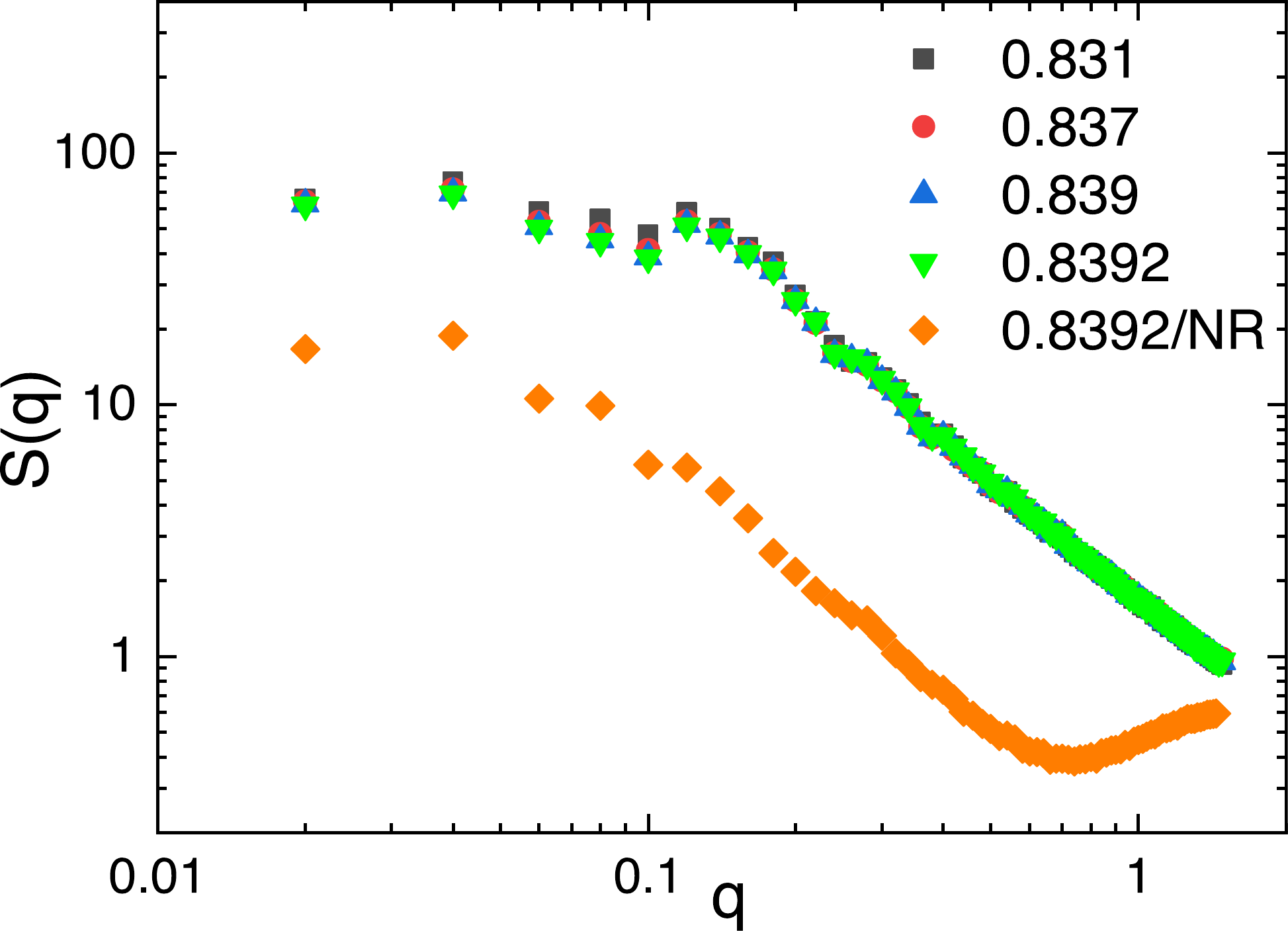}
\caption{$S(q)$ for {\color{black} the smaller of the two (by $N$)} $3d$ packings with $\beta=3.3$ and $\lambda= 50$ for the jammed configuration $(\rho = 0.8392$) and three earlier-stage configurations with volume fractions within 1\% of the jammed value. $S(q)$ calculated after removing rattlers is designated ``NR.'' }
\label{fig:S(q)-R-NR}
\end{figure}

{\color{black} The most computationally intensive part of producing jammed packings is the final stage when the residual kinetic energy of small-size, mechanically-unstable rattlers is quenched at nearly constant $\rho$.
Rattler motions are confined to pores defined by the stationary large non-rattler particles constituting the stable structural backbone.
Since the fractal behavior of $S(q)$ spans $2\pi/\lambda \lesssim q < 1$, estimates of $d_f$ should not depend on these still-mobile particles with size of order $\sigma$.
To test this point, we computed $S(q)$ as the $\beta = 3.3$ system approached jamming for three $\rho$ values within 1\% of the jammed volume fraction, $\rho = 0.8392$, where the last digit is given to differentiate the configurations.
As Fig.~\ref{fig:S(q)-R-NR} shows, $S(q)$ is unchanged at these smaller $\rho$, implying that small-size rattlers are irrelevant to fractal behavior provided that the larger-scale structure is in place.}

{\color{black} To further reinforce the notion that small-size rattlers do not influence $d_f$, they can be removed prior to calculating $S(q)$.
As indicated by $\phi_R$, this procedure removes significant fractions of the particle count for all power-law packings, with smaller impact on $\rho_{NR}$ (see Table~\ref{tab:table1}).
Figure~\ref{fig:S(q)-R-NR} shows the resulting $S(q)$ for $\beta = 3.3$ with rattlers removed, and illustrates that the fractal scaling is unaffected despite the removal of $91.1\%$ of all particles, accounting for $\rho - \rho_{NR} = 0.093$ of the particle volume fraction.
The reduction of $N$ is reflected in the downwards shift of the data, while the behavior of $S(q)$ for $q\sim 1$ implies that the loss of small particles does impact the structure over a range of high $q$ of order the inverse size of the rattlers.
This result supports the notion that the mechanically-stable backbone dictates packing fractal behavior. } 

In summary, packings of power-law-distributed $d$-dimensional spheres were created using DEM simulations.
This dynamics-based approach offers a counterpoint to packings generated through particle insertion which may not reflect physical processes and do not necessarily ensure mechanical stability.
A scheme was introduced for calculating static structure factors $S(q)$ for triclinic simulation cells; computed $S(q)$ for power-law packings were shown to be fractal.
For $2d$ packings, $d_f$ was weakly dependent on $\beta$ for $2.1 \le \beta \le 2.9$, while in $3d$, $d_f$ was independent of size distribution for $2.5 \le \beta \le 3.8 $ with constant value $d_f \approx 2.0$.
These results differ from recent findings of packings created via an RSA technique which found $d_f = \beta - 1$~\cite{cherny2023}.
Our work demonstrated that the observed fractal behavior was insensitive to whether the packing was fully or only nearly jammed.
Further, we showed that the removal of mechanically unstable particles before computing the structure factor did not affect $d_f$.
Both of these results stem from the packings deriving stability from the largest particles most of all \cite{monti2022a}, and from smaller particles to increasing degrees with increasing $\beta$, correlating with a decrease in the overall number fraction of mechanically unstable particles.

I.S. acknowledges support from the U.S. Department of Energy (DOE), Office of Science, Office of Advanced Scientific Computing Research, Applied Mathematics Program under Contract No. DE-AC02-05CH11231.
This work was performed in part at the Center for Integrated Nanotechnologies, a U.S. DOE and Office of Basic Energy Sciences user facility.
Sandia National Laboratories is a multimission laboratory managed and operated by National Technology \& Engineering Solutions of Sandia, LLC, a wholly owned subsidiary of Honeywell International Inc., for the U.S. DOE’s National Nuclear Security Administration under contract DE-NA0003525.
This paper describes objective technical results and analysis. 
Any subjective views or opinions that might be expressed in the paper do not necessarily represent the views of the U.S. DOE or the U.S. Government.

\appendix*
\section{Static structure factors for triclinic cells}
Static structure factors $S(q)$ are calculated as
\begin{equation}
    S(\mathbf{q}) = \frac{1}{N}\sum_{i,j=1}^{N}b_ib_je^{-i\mathbf{q}\cdot(\mathbf{r}_{i}-\mathbf{r}_{j})},
    \label{eqn:sq}
\end{equation}
where $b_i$ is the scattering length of particle $i$, and the summation is over all particle pairs in the unit cell.
Since we are interested in the packing fractal dimension, we set $b_i = 1$.
For cubic cells of length $L$, $\mathbf{q} = \frac{2\pi}{L}\mathbf{n}$, where $\mathbf{n}=\left[n_x,n_y,n_z\right]$ with $n_i = 0,\pm 1, \pm 2, ...$. 
For triclinic cells, calculation of $S(q)$ requires the associated $\mathbf{q}$. 
These emerge naturally if the triclinic cell is denoted by a transformation matrix $\boldsymbol{H}$, which relates the (untransformed) reduced coordinates $\mathbf{s}_{i}$ of particle $i$ within the domain to the transformed real-space coordinates $\mathbf{r}_{i}$ from the affine deformation of the periodic cell. 
These transformed coordinates are
\begin{equation}
    \mathbf{r}_{i} = \mathbf{L}_{1}s_{i,1} + \mathbf{L}_{2}s_{i,2} + \mathbf{L}_{3}s_{i,3} = \boldsymbol{H}\mathbf{s}_{i},
\end{equation}
where $\mathbf{s}_{i}=\left[s_{i,1},s_{i,2},s_{i,3}\right]$ are reduced-space coordinates such that $0\leq s_{i,\alpha} \leq 1$, and the cell vectors $\mathbf{L}_{1}$, $\mathbf{L}_{2}$, and $\mathbf{L}_{3}$ tile the space, such that their concatenation forms $\boldsymbol{H}$:
\begin{equation}
    \boldsymbol{H} = \begin{pmatrix} \mathbf{L}_{1} & \mathbf{L}_{2} & \mathbf{L}_{3} \end{pmatrix}.
\end{equation}

If $\boldsymbol{H}$ is upper triangular, as in LAMMPS~\cite{thompson2022}, it can be written as 
\begin{equation}
    \boldsymbol{H} = \begin{pmatrix} L_{x} & \alpha_{xy} & \alpha_{xz} \\0 & L_y & \alpha_{yz} \\0 & 0 & L_z \end{pmatrix},
\end{equation}
where $L_i$ represent the projected lengths of the triclinic cell along the Cartesian dimensions, and $\alpha_{ij}$ represent cell vector `tilts.'
Reciprocal vectors for the triclinic cell are obtained from the inverse of $\boldsymbol{H}$ as~\cite{nose1983}
\begin{equation}
    \boldsymbol{H}^{-1} = \begin{pmatrix} \mathbf{q}_{1}^{T} \\ \mathbf{q}_{2}^{T} \\ \mathbf{q}_{3}^{T} \end{pmatrix},
\end{equation}
where $T$ indicates the vector transpose.
For a given integer vector $\mathbf{n}$ specifying a periodic image, any $\mathbf{q}$ can be defined in terms of $\boldsymbol{H}$ as
\begin{equation}
    \mathbf{q} = 2\pi \boldsymbol{H}^{-T}\mathbf{n},
\end{equation}
with components:
\begin{align}
            q_x &= 2\pi \frac{n_x}{L_x}, \\
            q_y &= 2\pi \left(\frac{n_y}{L_y} - \frac{n_x \alpha_{xy}}{L_xL_y}\right), \\
            q_z &= 2\pi\left[ \frac{n_z}{L_z} - \frac{n_y \alpha_{yz}}{L_yL_z} + \frac{n_x (\alpha_{yz} \alpha_{xy} - L_y \alpha_{xz})}{L_xL_yL_z}\right].
\end{align}
The dot product $\mathbf{q}\cdot \mathbf{r}_{i}$ is independent of $\boldsymbol{H}$ and only depends on the reduced coordinates $\mathbf{s}_i$:
\begin{align*}
    \mathbf{q}\cdot \mathbf{r}_{i} &= 2 \pi \boldsymbol{H}^{-T}\mathbf{n} \cdot \boldsymbol{H}\mathbf{s}_i \\
                                   &= 2 \pi \mathbf{n}^{T} \mathbf{s}_i.
\end{align*}
$S(q)$ is computed by choosing $\mathbf{n}$ and then calculating the corresponding $\mathbf{q}=2\pi\boldsymbol{H}^{-T}\mathbf{n}$. 
Results for $S(q)$ are spherically averaged for $q = |\mathbf{q}|$. 

\bibliography{bib}

\end{document}